\newcommand\apj{{ApJ\,}}%
\newcommand\apjl{{ApJ\,}}%
\newcommand\apjs{{ApJS\,}}%
\newcommand\aap{{A\&A\,}}%
\newcommand\mnras{{MNRAS\,}}%
\documentclass[graybox, natbib, footinfo]{svmult}

\usepackage{mathptmx}       
\usepackage{helvet}         
\usepackage{courier}        
\usepackage{type1cm}        
\usepackage{makeidx}         
\usepackage{graphicx}        
\usepackage{multicol}        
\usepackage[bottom]{footmisc}

\makeindex             

\begin{document}

\title*{New Observational Constraints to Milky Way Chemodynamical models}
\author{Cristina Chiappini, Ivan Minchev, Friedrich Anders, Dorothee Brauer, Corrado Boeche, Marie Martig} 
\institute{Cristina Chiappini, Ivan Minchev, Friedrich Anders, Dorothee Brauer \at Leibniz-Institut f\"ur Astrophysik Potsdam (AIP), An der Sternwarte 16, 14482 Potsdam, Germany, \email{cristina.chiappini@aip.de}
\and Corrado Boeche {\at Astronomisches Rechen-Institut, Zentrum f\"ur Astronomie der Universit\"at Heidelberg, M\"onchhofstr. 12-14, D-69120 Heidelberg}
\and Marie Martig {\at Max-Planck-Institut f\"ur Astronomie, K\"onigstuhl 17, D-69177 Heidelberg, Germany}}
%
%
\maketitle

\abstract{Galactic Archaeology, i.e. the use of chemo-dynamical information for stellar samples covering large portions of the Milky Way to infer the dominant processes involved in its formation and evolution, is now a powerful method thanks to the large recently completed and ongoing spectroscopic surveys. It is now important to ask the right questions when analyzing and interpreting the information contained in these rich datasets. To this aim, we have developed a chemodynamical model for the Milky Way that provides quantitative predictions to be compared with the chemo-kinematical properties extracted from the stellar spectra. Three key parameters are needed to make the comparison between data and model predictions useful in order to advance in the field, namely: precise proper-motions, distances and ages. The uncertainties involved in the estimate of ages and distances for field stars are currently the main obstacles in the Galactic Archaeology method. Two important developments might change this situation in the near future: asteroseismology and the now launched Gaia. When combined with the large datasets from surveys like RAVE, SEGUE, LAMOST, Gaia-ESO, APOGEE , HERMES and the future 4MOST we will have the basic ingredients for the reconstruction of the MW history in hands. In the light of these observational advances, the development of detailed chemo-dynamical models tailored to the Milky Way is urgently needed in the field. Here we show the steps we have taken, both in terms of data analysis and modelling. The examples shown here illustrate how powerful can the Galactic Archaeology method become once ages and distances are known with better precision than what is currently feasible.}

\section{Galactic Archaeology: how powerful can it be?}
\label{sec:1}

The use of chemo-kinematical information of a large number of stars to understand which were the main steps in the formation of the Milky Way has been challenged when it became clear that stars can move away from their birthplaces. N-body simulations of the formation of disk galaxies have shown radial migration to be a common phenomena \cite[see][]{raboud98,sellwood02,roskar08a}. Also from the observational side, there have been
claims for detecting local stars that have exceedingly large metallicities to have been born in the Solar Neighbourhood (so-called super metal-rich stars). These stars are often old and show typical thin disk kinematics \cite[see][and references therein]{trevisan11}. As the metallicity of the interstellar medium (traced by HII regions and young stars near the Sun) is not expected to have increased much in the last 3-4 Gyrs \cite[see][]{chiappini03} due to a rather inefficient star formation rate at the solar Galactocentric radii during this period of time, these stars most probably came from the inner, more metal-rich, regions of the Galaxy. Indeed, chemical evolution models for the MW thin disk, which are compatible with the abundances measured in the interstellar medium at present time, cannot predict stars more metal rich than [Fe/H] $\sim$0.2~dex to be present at 8~kpc from the Galactic center \citep{chiappini09a}. 

The age and chemical composition of the stars constitute relic informations not disturbed by the radial migration process. However, pure chemical evolution models constrained by observations derived from stars currently in a certain volume (for instance, at the Solar vicinity) are in danger of  inferring wrong conclusions if the adopted observational samples turn out to have been severely contaminated by \emph{intruder} stars. Indeed, radial migration, if important, could affect classic constraints such as the age-metallicity relation, metallicity distributions, abundance ratios  vs. metallicity relations, as well as the radial abundance gradients. These stars cannot be identified by their kinematics because, as shown by \cite{sellwood02} radial migration means permanent changes to the stellar angular momenta, making it possible that a migrated star shows the exact same kinematics of the ones born in situ, in a certain studied volume.

In particular, radial migration\footnote{Which is different from the fact that in different samples there are some stars wandering from other regions due to eccentric orbits -- those can be easily identified by calculating their orbital parameters - see, for instance \citealt{anders14}).} has been used to challenge one of the main pillars of the two-infall model for the formation of the Milky Way \citep{chiappini97}. In \citet{chiappini97} we proposed two main epochs of star formation in the Galaxy, the first one related to the thick disk and the second one to the thin disk. This suggestion was based mostly on chemical evolution arguments (abundance ratios) given the impossibility of using ages as a constraint due to the large uncertainties at the time.  A low star formation period in between the two main infall episodes would naturally explain the observed \emph{gap} in the [$\alpha$/Fe] vs. [Fe/H] diagram \citep{fuhrmann98}. In \cite{chiappini09a} we computed a somewhat different model where the two infall episodes were completely disentangled and gas from the first infall did not pre-enrich the second component. One of the main conclusions of this model was that the chemical properties of the thick and thin disk (especially the shifts observed in several abundance ratios of stars kinematically classified as belonging to the thick or thin disk) can be well accounted for by a model were the thick disk formed on a short timescale ($\sim$1-2 Gyrs) and with a larger star formation efficiency than the thin disk.

However, the reality of the observed chemical discontinuity has been challenged by \cite{schonrich09b}, who constructed a model for chemical evolution of the MW, assuming a certain migration efficiency in order to fit the current  thin disk abundance gradient and metallicity distribution at the Solar vicinity. Contrary to \cite{chiappini97}, they find that radial migration would account for forming the thick disk, without the need of two main episodes of star formation. However, as discussed in \citep{mcm13}, in the absence of massive mergers disk thickening due to migration is insignificant: owing to the conservation of vertical action, only extreme migrators contribute by contracting the inner disk and thickening the outskirts. This deficiency in the Schoenrich and Binney model \cite{schonrich09b} affects their main conclusion that a thick disk can be formed in a merger-free MW disk evolution, and that the MW formation does not require two main episodes of star formation. In addition, N-body simulations where a bar is formed (as is the case in our Galaxy) suggest the efficiency of migration to vary with time and distance from the galactic center \citep{minchev10, brunetti11,shevchenko11}. This fact implies that analytical approaches are not suitable to properly investigate these issues. N-body simulations are the appropriate technique to treat time-dependent non-axisymmetric systems such as our barred galaxy. 

Apart from the inherent difficulties linked to the existence of radial migration and how to account for it in chemo-dynamical models, current samples of disk stars for which detailed chemical abundances and distances are known are still confined to a small volume of the Milky Way. In particular, there is always the danger that the studied stellar samples suffer from biases that could, in principle, artificially create discontinuities in the abundance ratio diagrams, as suggested by \citet{bovy12b}. While it is unclear whether the separation between the thin and thick disk in chemistry is real or due to selection effects in the SEGUE sample \citep{bovy12b}, or inexistent as in the RAVE samples (e.g. \citealt{boeche13a}), the fact that the unbiased volume-completed (although very local) sample of \cite{fuhrmann11} does show a gap argues that it may indeed be the legacy of two discrete stellar populations, or star formation episodes. Recently, \citet{anders14}, using a sample of red giants from the first year of APOGEE data, also found a clear discontinuity in the chemical plane. As discussed in that paper, given the way this sample was selected, it was very unlike that the observed discontinuity would be produced by sample selection biases. This was now recently confirmed by Nidever et al. [submitted]  who has used a sample of APOGEE red clump stars, for which the sample selection was taken into account, and still report similar {\it gaps} as the ones seen in the \citet{anders14} paper.  In these studies, one of the main uncertainties remaining is the proper motions, but this should be improved soon now with the launched Gaia satellite.

Finally, an important role can be played by asteroseismology as well. Indeed, as discussed during this meeting, solar-like pulsating red giants offer a well-populate class of objects, not only spanning a large age range, but also large distances\footnote{Even though, only in the particular fields observed by CoRoT and {\it Kepler}. Hopefully, with Plato much the same data can be obtained for a larger portion of the Galaxy -- see \cite{plato14c}}, for which is now in principle possible to obtain accurate distance and ages \cite[Valentini, Miglio, this conference, and][]{miglio13}.

Given the suggestions in the Literature (mostly made in 2008/2009) that a) the observed chemical discontinuity in the solar vicinity stars was due to selection biases, and b) that radial migration could account for the observations of thick disk stars without the need for a two-infall model, we decided to proceed in the following two directions:

\begin{itemize}
\item develop a chemodynamical model (within the cosmological framework) and try to understand what is possible in terms of thick disk formation and if one simple disk (a single star formation episode model) can reproduce the chemo-kinematical relations in the MW.
\item investigate the chemo-kinematic properties of large stellar samples in the MW, for which biases can be estimated (using data from RAVE, SEGUE and APOGEE surveys); with the goal to find new, more tight constraints to models for the MW.
\end{itemize}

In the last couple of years we have progressed in both directions and some of the results obtained so far are described in the next Sections.

\section{The MCM Chemodynamical model for the Milky Way}
\label{sec:2}

We have recently developed an alternative way to construct a chemo-dynamical model for our Galaxy, here referred as the MCM model \citep{mcm13}, which is the fusion between a state-of-the-art simulation\footnote{The simulation builds up a galactic disk self-consistently by gas inflow from filaments and mergers and naturally takes into account radial migration processes due to early merger activity and internal disk evolution at low redshift. A central bar is developed early on, similar in size at the final simulation time to that of the MW.} in the cosmological context \citep{martig12} and a detailed thin-disk chemical evolution model. The key point of our approach is the use of the exact star formation history and chemical enrichment from our chemical model to assign to the particles of the simulation. This novel approach was born from the need to avoid the known problems with chemical enrichment and star formation currently found in fully self-consistent simulations (see discussion in \citealt{mcm13}), and to focus on quantifying the radial migration and its impact on the classic MW constraints such as metallicity distribution functions (MDFs), age-metallicity relation and abundance gradients. 

\begin{figure}[h]
\center
\includegraphics[scale=.5]{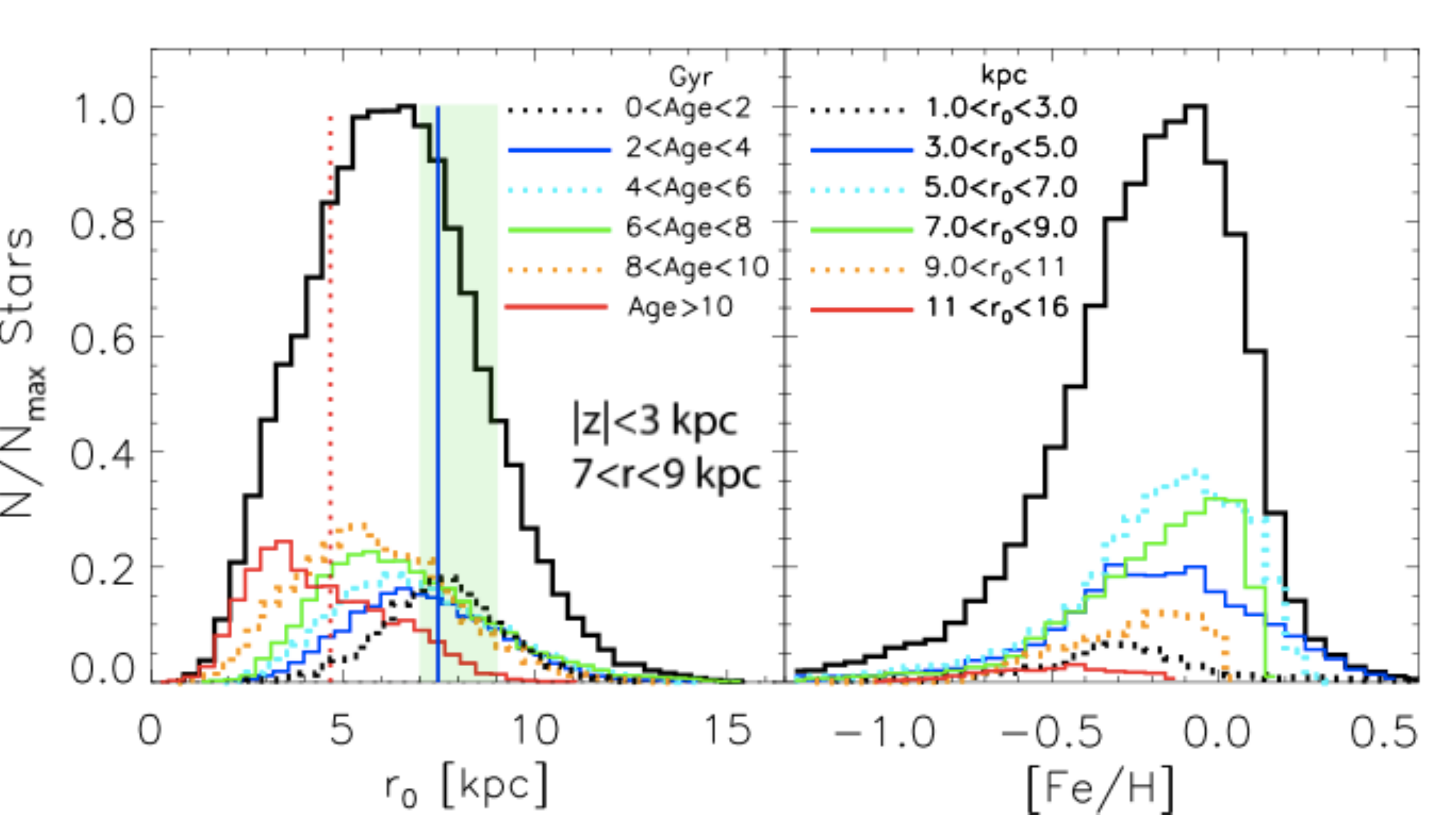}
\caption{For stars currently at the Solar vicinity we show: a) one the left their age and birth radius distributions, and b) on the right their metallicity distributions (the solid black line refers to the the total MDF). Figure taken from \cite{mcm13}. }
\label{fig1}       
\end{figure}

\begin{figure}[h]
\center
\includegraphics[scale=.28]{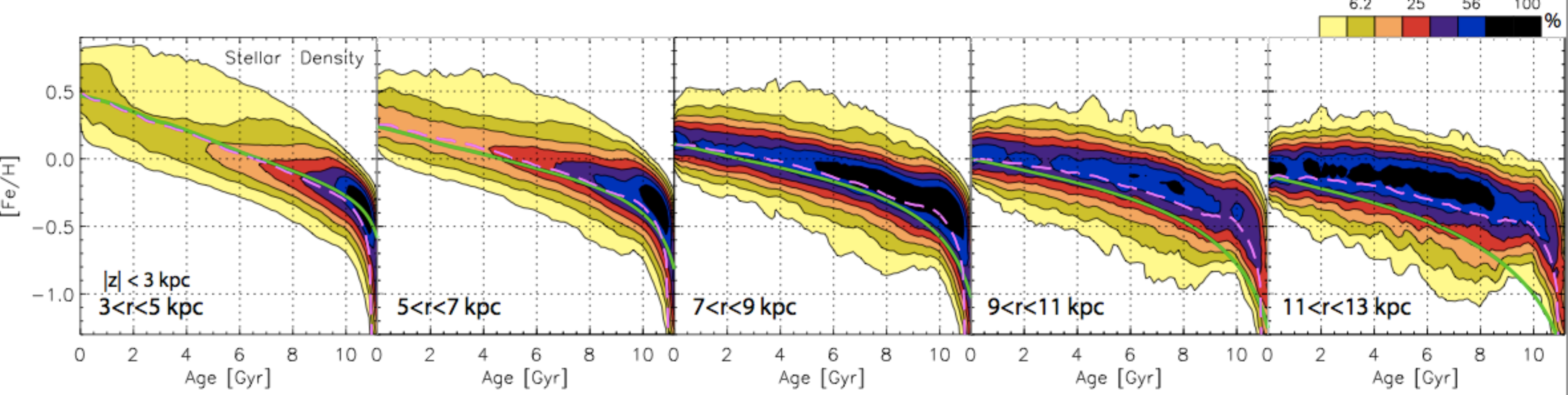}
\caption{The Age-metallicity relation predicted at different radial bins. The color code indicates how the stellar density varies in each diagram. The middle row (7$ < r <$ 9~kpc) corresponds to the solar neighborhood. The input chemistry, native to each bin, is shown by the solid-green curve. The dashed-pink curve indicates the mean
[Fe/H] binned by age which takes into account radial migration effects. Figure adapted from paper II. }
\label{fig2}       
\end{figure}

The main result of the MCM model for the Solar vicinity is shown in Fig.~\ref{fig1}. The model predictions imply that stars currently at the Solar vicinity (here defined as stars with galactocentric distances 7$<$ R $<$9 kpc and height above the plane $z < $3 kpc)  are a mosaic of stars born at different locations ($r_0$) and at different times, with a peak at $r_0 \sim$6~kpc. We also predict more than 60\% of stars currently at the 7 $<$ R $<$ 9 kpc bin to come from inner regions, whereas only around 10\% have birth radii beyond the solar circle. Because the chemical evolution at a distance of 6~kpc from the galactic center does not differ much form that at 8~kpc, the impact of radial migration on the Solar vicinity is minor. This can be seen in Fig.~\ref{fig2} where the predicted age-metallicity relation for stars in the Solar circle is shown, together with the same prediction for other galactocentric bins.
We find in our Paper II \citep{minchev14a} that the width of the $r_0$-distribution increases between the innermost and outermost annuli considered. The predicted increase in contamination from migration with radius is due to the exponential drop in disk surface density, where, on the average, stars migrate larger distances outwards than inwards. Inward migration is still very important, and has the effect of balancing the contribution from stars coming from the inner disk, at intermediate radii. 

\begin{figure}[h]
\center
\includegraphics[scale=.24]{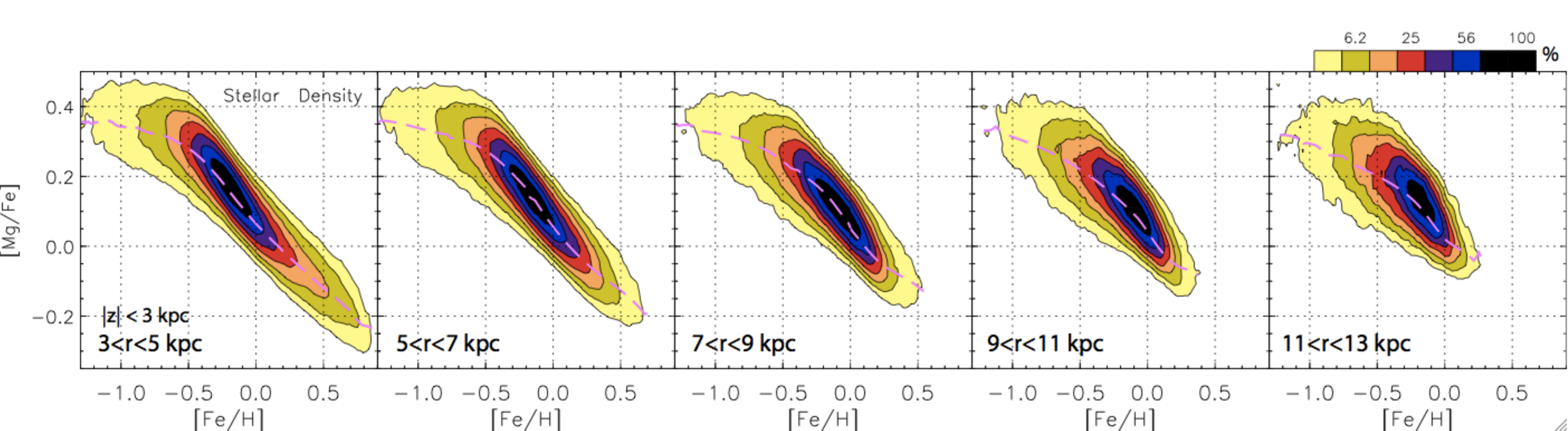}
\caption{The predicted density distribution of stars in a [Mg/Fe] vs. [Fe/H] diagram, at different radial bins. The dashed-pink curve indicates the mean [Mg/Fe]
binned by [Fe/H]. The model, predict no gap in the diagram, even though several properties of its oldest stars do match several typical thick-disk kinematical, structural and chemical properties. Figure adapted from Paper II.}
\label{fig3}       
\end{figure}

In Fig.~\ref{fig3} the same is shown for the predicted density distribution of stars in the [Mg/Fe] vs. [Fe/H] diagram. Notice that the chemo-dynamical model of a pure thin disk does not predict a gap in this diagram. On the other hand, all the other properties normally attributed to thick disk stars are met by the oldest (ages $>$ 10 Gyr)  stars in our simulation. At the Solar vicinity the oldest stars show typical thick disk kinematical (as for instance, rotation velocity lag of thing disk stars, larger velocity dispersion), structural (such as shorter scale-length than younger stars), and chemical properties (such as the metallicity and [Mg/Fe] distributions). These surprising results may indicate that a discrete thick-disk component (in time, and with a high-efficiency star formation rate - as suggested in \citealt{chiappini09a}) might be still needed, which we will investigate in the near future. In other words, the discontinuity in the chemical diagram seems to be really an effect of different star formation regimes.

\begin{figure}[h]
\center
\includegraphics[scale=.35]{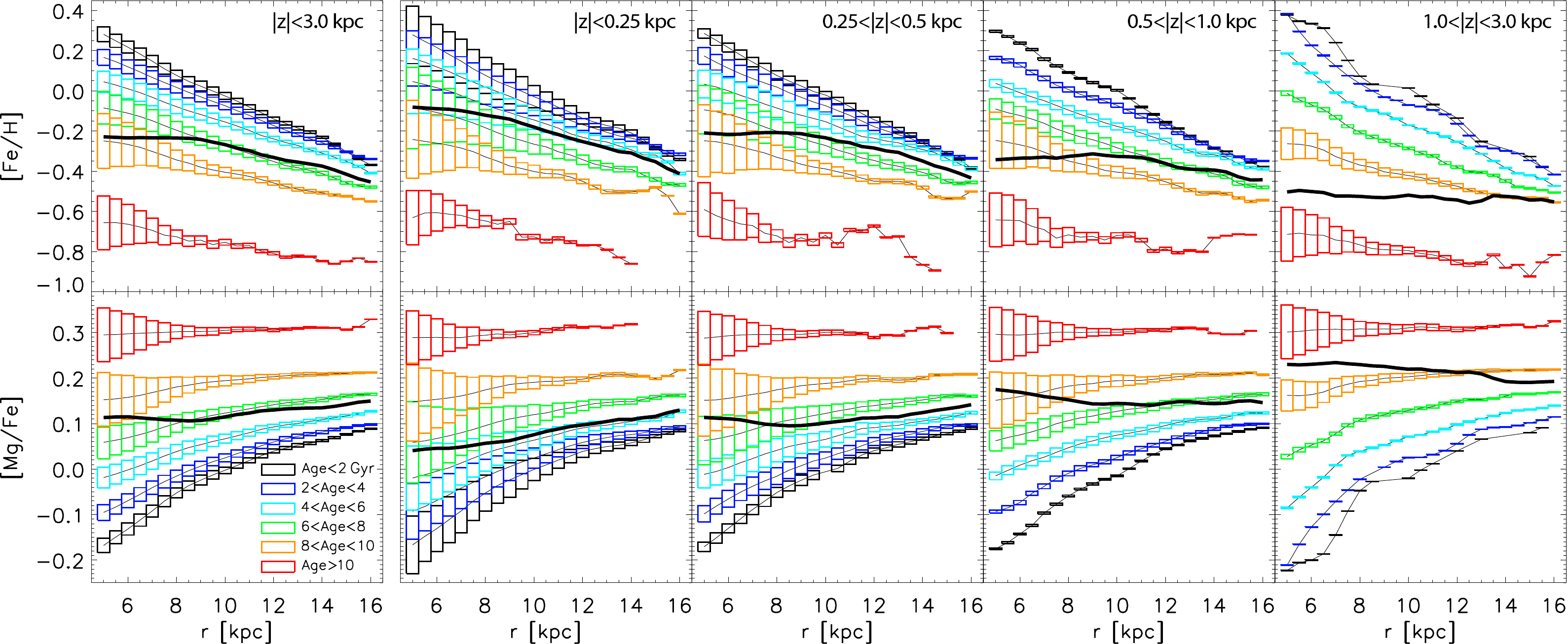}
\caption{The predicted density distribution of stars in a [Mg/Fe] vs. [Fe/H] diagram, at different radial bins. The dashed-pink curve indicates the mean [Mg/Fe]
binned by [Fe/H]. The model, predict no gap in the diagram, even though several properties of its oldest stars do match several typical thick-disk kinematical, structural and chemical properties. Figure adapted from Paper II.}
\label{fig4}       
\end{figure}

In our paper II \cite{minchev14a} we also explain the variation of observed abundance gradients with distance from the mid-plane as due to a different mix of stars of different ages at the different slices in $z$. This is very important as studies using different samples (see next Section) have found different metallicity and [$\alpha$/Fe] gradients in the MW. Our main predictions are shown in Fig.~\ref{fig4}. The thick black curves in the top row show the azimuthally averaged metallicity variation with galactic radius for stellar samples at different distances from the disk midplane, as marked in each panel. Different colors correspond to different age groups as indicated in the bottom-left panel. The height of rectangular symbols reflects the density of each bin. The bottom row of Fig.~\ref{fig4} shows the same information as above but for [Mg/Fe].

Other predictions of the MCM model can be seen in Paper I \citep{mcm13} and Paper II \citep{minchev14a}.

\section{New constraints from RAVE, SEGUE and APOGEE}
\label{sec:3}
More detailed constraints, as for instance, sampling different heights from the galactic plane, do play a key role in constraining theoretical models.
Here I give some examples of recent new constraints to chemodynamical modes we have obtained by analyzing RAVE, SEGUE and APOGEE samples. New constraints are also now coming from the Gaia-ESO survey (see contributions in this volume).

\subsection{Chemodynamics with RAVE}

By studying a sample of giants, we \citep{boeche13a} have shown that despite RAVE's modest R=7000 resolution, the inferred chemical abundances and kinematical parameters can be confidently used to investigate chemodynamical properties of stars in a z$_{max}$-eccentricity plane (where z$_{max}$ is the maximum height from the Galactic plane in the star's orbit).  It turns out that different portions of such a diagram are more or less affected by migrators. By comparing the chemodynamical model predictions with diagrams of this kind, it is possible to look for the best parameter space in which to detect radial migrators, based on chemistry and age knowledge (hopefully feasible in the near future with CoRoT and {\it Kepler} data).

More recently, we discovered \citep{minchev14b}, that the velocity dispersion of a sample of RAVE giant stars decreases strongly for stars with large [Mg/Fe]  (the oldest stars). This findings, although at odd with the classical expectations that older populations should show larger velocity dispersions, can be understood in the following way:  perturbations from massive mergers in the early universe do affect more strongly the outer parts of the disk and, at the same time, lead to the subsequent radial migration of stars with cooler kinematics from the inner disk. Similar reversed trends in velocity dispersion are also found for different metallicity subpopulations. Our results suggest that the Milky Way disk merger history can be recovered by relating the observed chemo-kinematic relations to the properties of past merger events. 

\begin{figure}[h]
\center
\includegraphics[scale=.4]{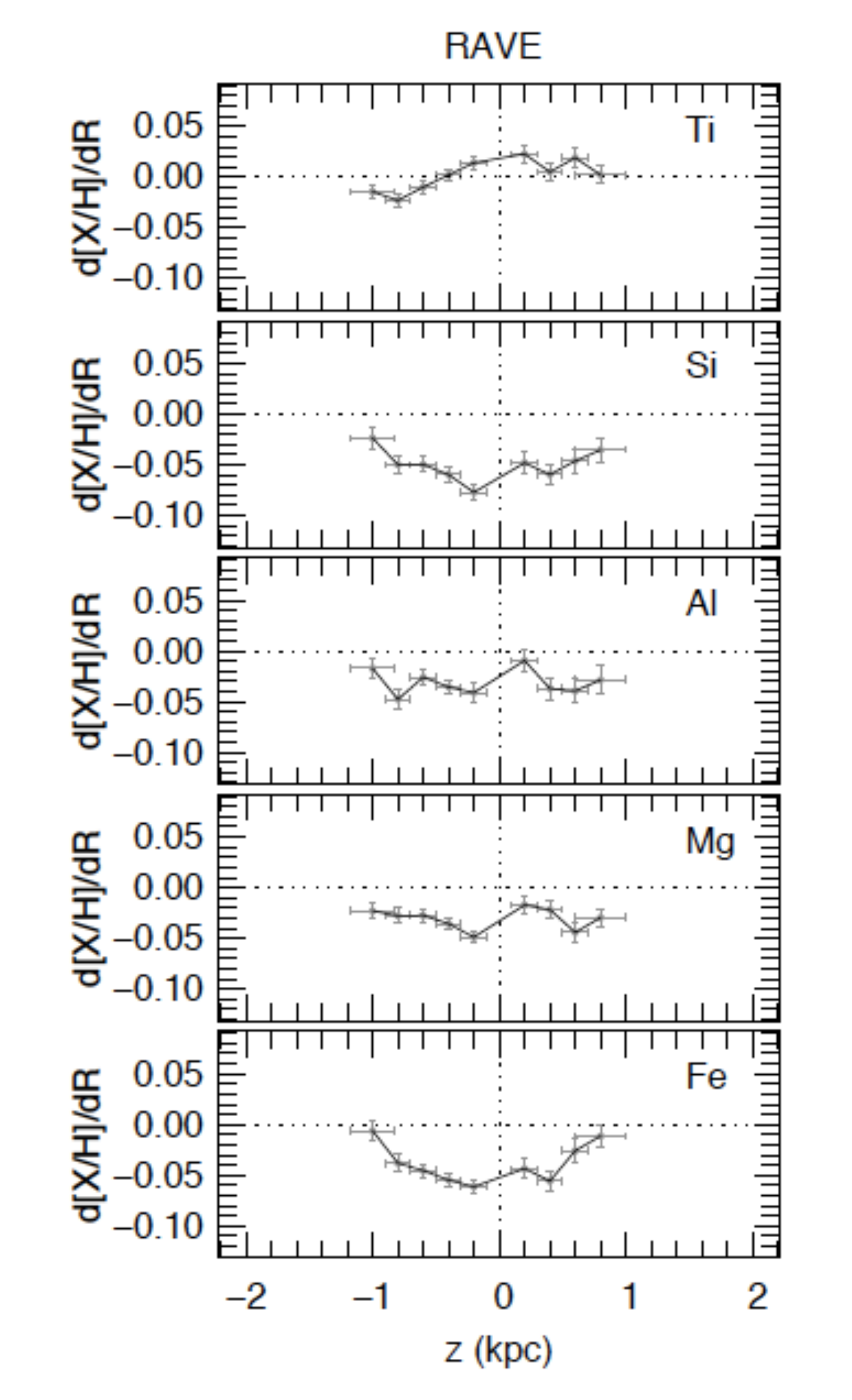}
\includegraphics[scale=.4]{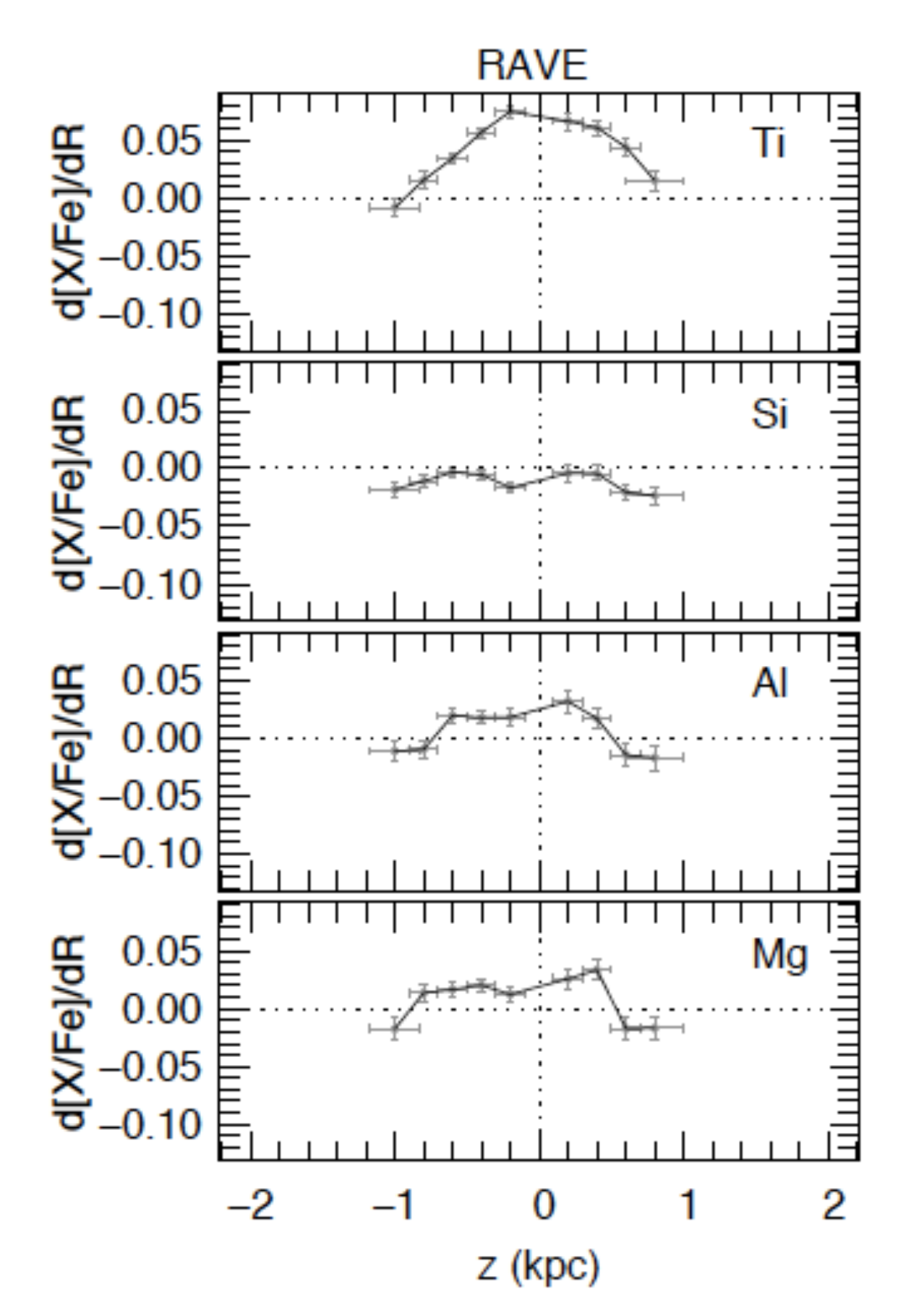}
\caption{Radial gradients for the RAVE red clump sample as a function of the distance from the
Galactic plane z (kpc) for the elemental abundances (left panel) and abundance ratios [X/Fe] (right panel). Figure adapted from \cite{boeche14}).}
\label{fig5}       
\end{figure}

Finally in \citealt{boeche13b,boeche14} we computed the abundance gradients for Mg, Al, Si, Ti, and Fe at different distances from the middle plane, using RAVE samples of dwarf and giant stars, respectively (see Fig.~\ref{fig5}). This was the first time abundance gradients for individual chemical elements have been studied in such large samples, and covering such large regions of the galactic disk. We find that a) the radial chemical gradients are negative and become progressively
flatter with increasing distance from the mid-plane; b) the vertical chemical gradients are negative and become progressively steeper as the distance from the plane increases, until it flattens out, c)  the high [$\alpha$/Fe] stars have radial chemical gradients consistent with zero; d) the vertical chemical gradients of the low [$\alpha$/Fe] stars are consistent with zero or with being slightly negative; e) the vertical chemical gradients of the high [$\alpha$/Fe] stars are consistent with zero, and negative once one approaches the middle plane. These general properties are matched by the MCM model, but a detailed comparison via mock-catalogues has to be carry out to quantitatively better constrain such models.

\subsection{Chemodynamics with the 1st year of APOGEE data}

More constraints were recently obtained from the study of a sample of giant stars observed by the Apache Point Observatory Galactic Evolution Experiment (APOGEE). APOGEE is the first multi-object high-resolution fiber spectrograph
in the near-infrared and hence is able to peer through the dust that obscures stars in the Galactic disc and bulge in the shorter wavelengths. Therefore the APOGEE data greatly complements studies such as RAVE and SEGUE which do not cover much the mid-plane. We selected a high-quality sample in terms of chemistry (amounting to around 20 000 stars) from the first year of APOGEE data and computed distances and orbital parameters for this sample, in order to formulate constraints on Galactic chemical and chemo-dynamical evolution processes in the solar neighborhood and beyond (e.g., metallicity distributions MDFs, [$\alpha$/Fe] vs. [Fe/H] diagrams, and abundance gradients). 

\begin{figure}[h]
\center
\includegraphics[scale=.35]{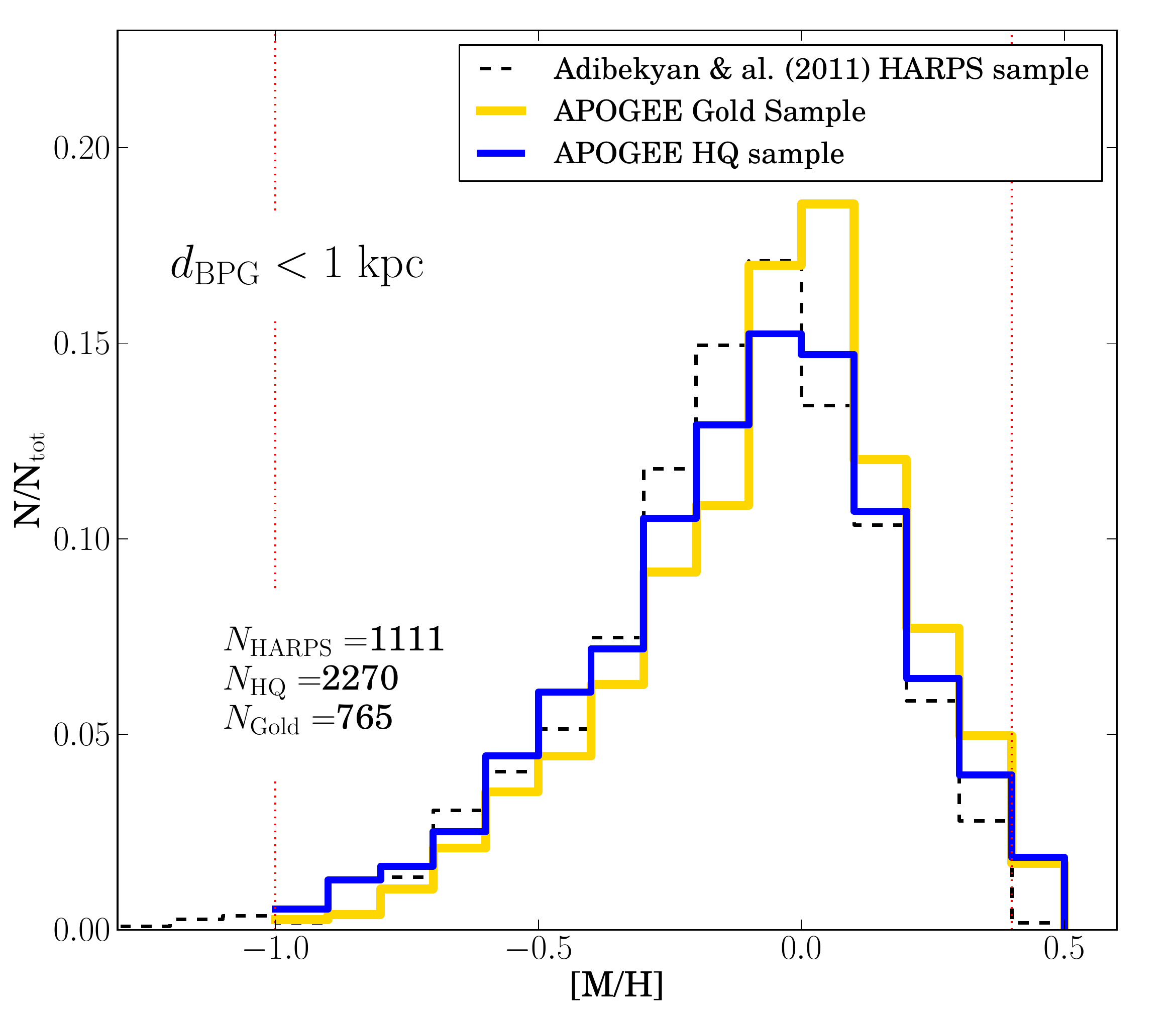}
\caption{The local MDF peaks slightly below solar metallicity both in the our APOGEE samples (the gold sample is a sub-sample with the best distances and proper motions) and in the HARPS sample. All three MDFs exhibit an extended tail towards [Fe/H] $= -$1, while showing a sharper cutoff at larger metallcities (the APOGEE sample shows a slight overabundance of stars with metallicities larger than around 0.3 dex with respect to what is observed in the HARPS sample.}
\label{fig6}       
\end{figure}

Our sample of red giant stars covers a heliocentric distance range of 10 kpc (most of the stars are situated 1-6 kpc away from the Sun), which enabled us to increase the Galactic volume studied with spectroscopic  stellar surveys with respect to the most recent chemodynamical studies based on RAVE and SEGUE  by at least a factor of 8. We find excellent agreement between the MDF of the local (d $<$ 100 pc) high-resolution high-S/N HARPS sample of \cite{adibekyan11} and our local APOGEE sample (with d $ <$ 1 kpc; see Fig.~\ref{fig6}). 

\begin{figure}[h]
\center
\includegraphics[scale=.3]{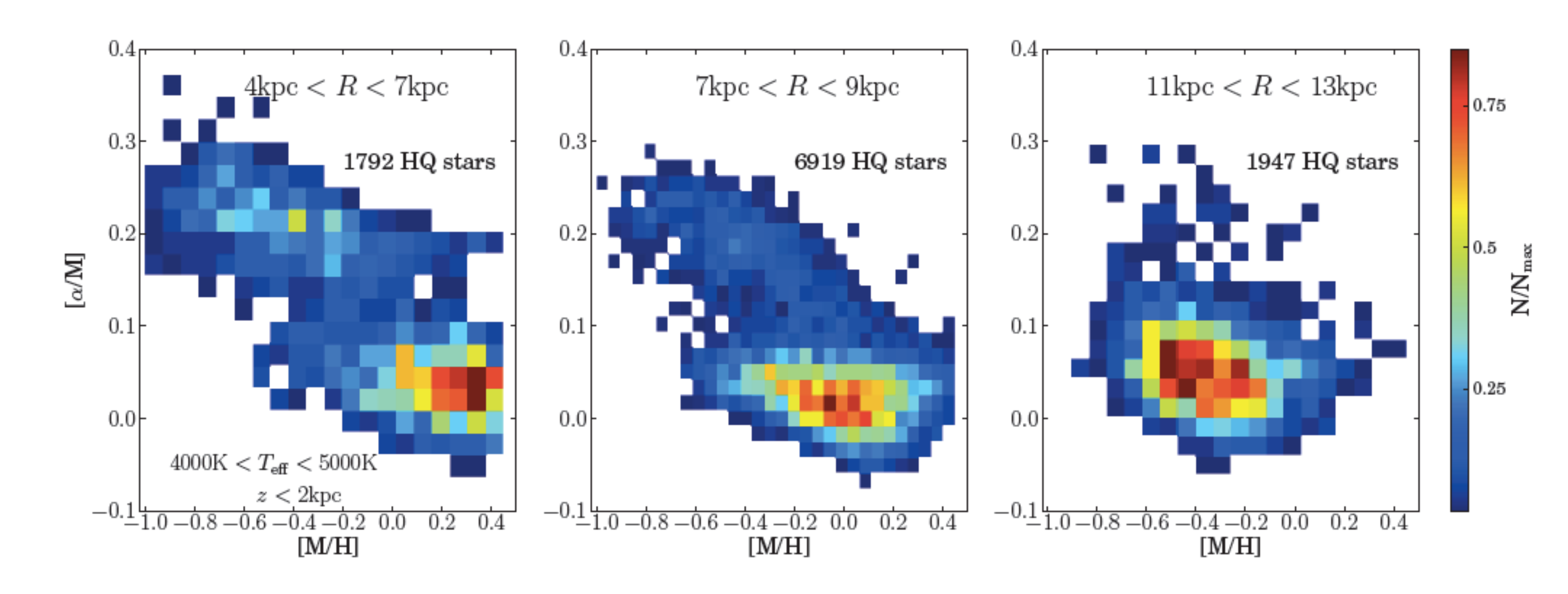}
\caption{Density plot of the chemical abundance plane in different radial bins for APOGEE red giant stars in the temperature range of 4000 K $<$ Te $<$ 5000 K and within z = 2 kpc. We confirm the result of \cite{bensby11} that
the radial scale length of the thick disc is much shorter than that of the thin disc. Also important is to notice that also in the inner region there is a {\it gap} in the chemical plane, as observed also for the solar neighborhood.}
\label{fig7}       
\end{figure}

More importantly, both samples compare well in an [$\alpha$/Fe] vs. [Fe/H] diagram, and in both cases a clear {\it gap} is observed.  Figure~\ref{fig7} shows this chemical diagram for three different Galactocentric distance bins (inner disk, solar vicinity and outer disk). We find $\alpha$-enhanced stars to be extremely rare in the outer Galactic disc, as has been suggested in the literature (e.g. \citealt{bensby11}).
We have also measured gradients in [Fe/H] and [$\alpha$/Fe] and their respective distribution functions over a range of 6 $<$ R $< $11 kpc in Galactocentric distance, and a 3 kpc range of distance from the Galactic plane, which are in fair agreement with the gradients derived from the SEGUE, GCS and RAVE samples. For stars located at 1.5 $<$ z$ <$ 3 kpc (which were not present in previous samples), we find a slightly positive metallicity gradient and a negative gradient in [$\alpha$/Fe]. 

\subsection{Chemodynamics with SEGUE}

We have selected a sample of G-dwarf stars from the SEGUE survey Data Release 9 \citep{dr9}. This sample (limited to distances below 3 kpc  from us to keep the uncertainties in distance and proper motion at a minimum) is complementary to the RAVE giant sample discussed in \cite{boeche13a}: both samples cover almost the same volume, whereas RAVE and SEGUE cover the southern and northern hemisphere, respectively. Details of our work can be found in Brauer et al. (in preparation). 

\begin{figure}[h]
\center
\includegraphics[scale=.55]{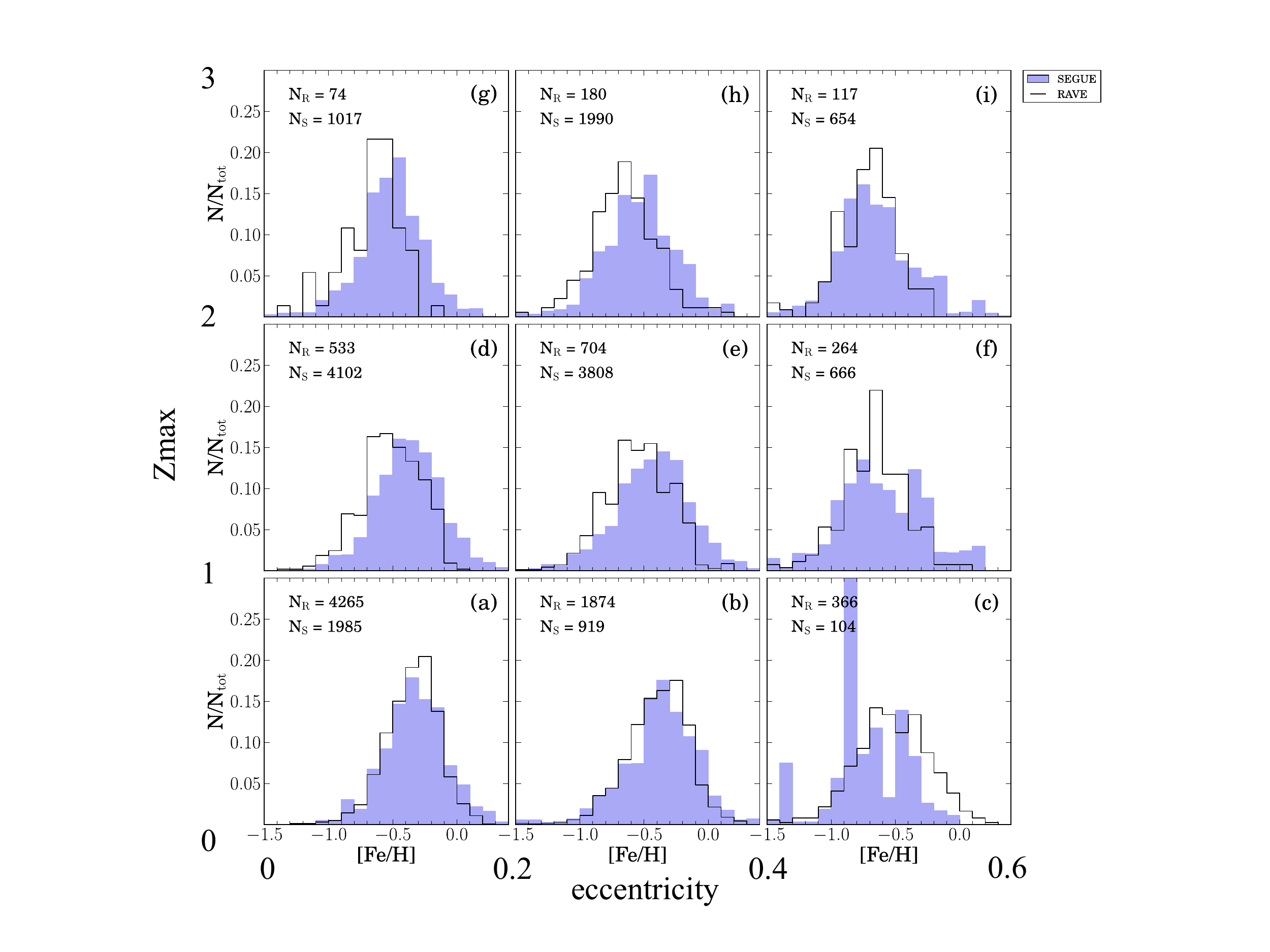}
\caption{This figure shows the metallicity distribution functions for a sample of RAVE giants (solid black histograms) and SEGUE dwarfs (filled blue histograms) at different ranges of z$_{max}$ and eccentricity (indicated by the larger numbers in the x- and y-axis). The number of objects, both for RAVE and SEGUE, in each of the z$_{max}$-eccentricity bins are also indicated. Diagrams such as this one will also be produced for our chemodynamical models as well as for the APOGEE sample.}
\label{fig8}       
\end{figure}

Fig.~\ref{fig8} shows the metallicity distributions obtained in both cases, RAVE and SEGUE, at the different locus of the z$_{max}$-eccentricity plane \citep{boeche13a} (the  z$_{max}$ and eccentricity bins are indicated in the figure). In the case of the SEGUE sample selection biases, not affecting the RAVE sample, had to be taken into account to allow a proper comparison of both surveys. Once this is done the agreement between the two samples within the different {\it orbital families} is encouraging. Indeed, these are completely different spectroscopic surveys, not only sampling a different hemisphere, but with different wavelength range, resolution, and different pipelines (moreover, we are here comparing a sample of dwarfs with a sample of giants). The agreement is particular good for the local thin disk dominated samples (stars in more circular orbits, not going above 1 kpc from the galactic plane - bottom left panel). More details can soon be found in Brauer et al. (in preparation).

\section{Conclusions and Outlook}
\label{sec:4}
Our main conclusions are summarized below:
\begin{itemize}
\item Stars observed ÒhereÓ are combination of stars born hot plus radial migration bringing contribution of old stars from inner radii;
\item The MW Òthick discÓ emerges naturally from (i) stars born with high velocity dispersions at high redshift, (ii) stars migrating from the inner disk early on due to strong merger activity, and (iii) further radial migration driven by the bar and spirals at later times;
\item The oldest population in our chemodynamical model has the properties of what has been called Òthick diskÓ, but not a clear gap in chemical plane. 
When applying to the simulated particles similar Òobservational biasesÓ as it is the case in some of the high-resolution observed samples, one can recover the chemical discontinuity. However, there are samples that seem to be free from biases and still show a gap (as for instance the APOGEE sample studied in \citealt{anders14}). This might indicate the need for the two star formation episodes as in the model of \cite{chiappini09a};
\item For now we are using [$\alpha$/Fe] as proxy for age. Tighter constraints will be obtained once ages will be available;
\item New observational constraints to theoretical models are now available from surveys like SEGUE, APOGEE, RAVE (some of the most recent summarized here) and Gaia-ESO. However, for a proper comparison with chemodynamical models one has to build {\it observed theoretical samples} (or mock catalogues). We plan to employ a newly developed selection interface (Piffl et al. in prep.) to
create mock surveys from a full chemo-dynamical MW model to be able to better constrain the models;
\item New constraints involving age knowledge are also now feasible thanks to the recent spectroscopic follow-ups of CoRoT and {\it Kepler} targets by both APOGEE and Gaia-ESO. These will be  presented in forthcoming papers (Anders et al., Martig et al. and  Valentini et al. in preparation).

\end{itemize}

\begin{acknowledgement}
C. C. would like to thank SDSS-III (and in particular the Brazilian and German Participation Groups) for the work done in APOGEE and SEGUE, and in particular Basilio Santiago and Leo Girardi. The RAVE collaboration is greatly acknowledged as well. Finally, I would like to thank the organizers for the invitation and for the patience in waiting for this contribution.
\end{acknowledgement}
%


\end{document}